\documentclass[magazine]{IEEEtran}
\usepackage{url}
\usepackage[utf8]{inputenc}
\DeclareUnicodeCharacter{00B2}{\textsuperscript{2}}
\usepackage{xcolor}
\usepackage{amsmath}

\usepackage{multirow}
\usepackage{amssymb}
\usepackage{cite}
\usepackage{enumitem}
\usepackage{bm}
\usepackage{graphicx}
\usepackage{amsthm}

\usepackage{comment}

\usepackage{gensymb}

\usepackage[acronyms,nonumberlist,nopostdot,nomain,nogroupskip]{glossaries}
\usepackage{tablefootnote}
\usepackage{booktabs}
\usepackage{tabularx}
\usepackage{epsfig}
\usepackage{soul}
\usepackage{xcolor}
\sethlcolor{yellow}
\usepackage{tikz}
\usetikzlibrary{external}
\usepackage{pgfplots}
\pgfplotsset{compat=newest} 
\pgfplotsset{plot coordinates/math parser=false}

\usetikzlibrary{plotmarks,patterns,patterns.meta,decorations.pathreplacing,backgrounds,calc,arrows,arrows.meta,spy,matrix}
\usepgfplotslibrary{patchplots,groupplots}
\usepackage{tikzscale}
\usepackage{siunitx}
\usepackage{tablefootnote}

\usepackage{multirow}
\usepackage{algorithm2e, setspace}
\SetAlCapNameFnt{\footnotesize}
\SetAlCapFnt{\footnotesize}
\RestyleAlgo{ruled}
\SetKwComment{Comment}{/* }{ */}
\usepackage{algpseudocode}

\usepackage[font=scriptsize]{subcaption}
\usepackage[font=footnotesize]{caption}

\usepackage{mathtools}

\usepackage{dblfloatfix}    % To enable figures at the bottom of page
\usepackage{colortbl}
\usepackage{footnote}

\usepackage{lipsum,afterpage,refcount}

\newacronym{3gpp}{3GPP}{3rd Generation Partnership Project}
\newacronym{adc}{ADC}{Analog to Digital Converter}
\newacronym{5g}{5G}{5th generation}
\newacronym{6g}{6G}{6th generation}
\newacronym{ai}{AI}{Artificial Intelligence}
\newacronym{aimd}{AIMD}{Additive Increase Multiplicative Decrease}
\newacronym{am}{AM}{Acknowledged Mode}
\newacronym{amc}{AMC}{Adaptive Modulation and Coding}
\newacronym{aqm}{AQM}{Active Queue Management}
\newacronym{awgn}{AGWN}{Additive White Gaussian Noise}
\newacronym{balia}{BALIA}{Balanced Link Adaptation}
\newacronym{bdp}{BDP}{Bandwidth-Delay Product}
\newacronym{bf}{BF}{beamforming}
\newacronym{cc}{CC}{Congestion Control}
\newacronym{cdf}{CDF}{Cumulative Distribution Function}
\newacronym{cn}{CN}{Core Network}
\newacronym{cqi}{CQI}{Channel Quality Information}
\newacronym{cp}{CP}{Control Plane}
\newacronym{csirs}{CSI-RS}{Channel State Information - Reference Signal}
\newacronym{dc}{DC}{Dual Connectivity}
\newacronym{rb}{RB}{Resource Block}
\newacronym{dce}{DCE}{Direct Code Execution}
\newacronym{dci}{DCI}{Downlink Control Information}
\newacronym{udp}{UDP}{User Datagram Protocol}
\newacronym{dl}{DL}{downlink}
\newacronym{fcfs}{FCFS}{first-come-first-served}
\newacronym{dmr}{DMR}{Deadline Miss Ratio}
\newacronym{fspl}{FSPL}{free-space path loss}
\newacronym{dmrs}{DMRS}{DeModulation Reference Signal}
\newacronym{e2e}{E2E}{End-to-End}
\newacronym{ppp}{PPP}{Poission Point Process}
\newacronym{aoi}{AoI}{Area of Interest}
\newacronym{cpu}{CPU}{Central Processing Unit}
 \newacronym{gpu}{GPU}{Graphics Processing Unit}
 \newacronym{tpu}{TPU}{Tensor Processing Unit}
\newacronym{si}{SI}{Study Item}
\newacronym{ecn}{ECN}{Explicit Congestion Notification}
\newacronym{edf}{EDF}{Earliest Deadline First}
\newacronym{enb}{eNB}{eNodeB}
\newacronym{epc}{EPC}{Evolved Packet Core}
\newacronym{es}{ES}{Edge Server}
\newacronym{cav}{CAV}{Connected and Autonomous Vehicle}
\newacronym{fdma}{FDMA}{Frequency Division Multiple Access}
\newacronym{fdd}{FDD}{Frequency Division Duplexing}
\newacronym{upa}{UPA}{Uniform Planar Array}
\newacronym{car}{CAR}{Circular Aperture Reflector }
\newacronym[firstplural=Radio Access Technologies (RATs)]{rat}{RAT}{Radio Access Technology}
\newacronym[firstplural=Radio Access Technology (RTs)]{rt}{RT}{Radio Technology}
\newacronym{fs}{FS}{Fast Switching}
\newacronym{isd}{ISD}{inter-site distance}
\newacronym{ftp}{FTP}{File Transfer Protocol}
\newacronym{gnb}{gNB}{Next Generation Node Base}
\newacronym{harq}{HARQ}{Hybrid Automatic Repeat reQuest}
\newacronym{hetnet}{HetNet}{Heterogeneous Network}
\newacronym{hh}{HH}{Hard Handover}
\newacronym{hol}{HOL}{Head-of-Line}
\newacronym{ia}{IA}{Initial Access}
\newacronym{imt}{IMT}{International Mobile Telecommunication}
\newacronym{iot}{IoT}{Internet of Things}
\newacronym{los}{LOS}{Line of Sight}
\newacronym{lte}{LTE}{Long Term Evolution}
\newacronym{m2m}{M2M}{Machine to Machine}
\newacronym{mac}{MAC}{Medium Access Control}
\newacronym{mc}{MC}{Multi-Connectivity}
\newacronym{mcs}{MCS}{Modulation and Coding Scheme}
\newacronym{mec}{MEC}{Mobile Edge Cloud}
\newacronym{mi}{MI}{Mutual Information}
\newacronym{mimo}{MIMO}{Multiple Input Multiple Output}
\newacronym{mmwave}{mmWave}{millimeter wave}
\newacronym{mptcp}{MPTCP}{Multipath TCP}
\newacronym{mr}{MR}{Maximum Rate}
\newacronym{mss}{MSS}{Maximum Segment Size}
\newacronym{mtd}{MTD}{Machine-Type Device}
\newacronym{mtu}{MTU}{Maximum Transmission Unit}
\newacronym{nfv}{NFV}{Network Function Virtualization}
\newacronym{vnf}{VNF}{Virtualization Network Function}
\newacronym{gv}{GV}{ground vehicle}
\newacronym{gvs}{GVs}{ground vehicles}
\newacronym{vec}{VEC}{Vehicular Edge Computing}
\newacronym{sdn}{SDN}{Software Defined Networking}
\newacronym{nlos}{NLOS}{Non Line of Sight}
\newacronym{nlosb}{NLOSb}{Building Non Line of Sight}
\newacronym{nlosv}{NLOSv}{Vehicle Non Line of Sight}
\newacronym{nr}{NR}{New Radio}
\newacronym{ofdm}{OFDM}{Orthogonal Frequency Division Multiplexing}
\newacronym{pdcch}{PDCCH}{Physical Downlonk Control Channel}
\newacronym{pdcp}{PDCP}{Packet Data Convergence Protocol}
\newacronym{pdsch}{PDSCH}{Physical Downlink Shared Channel}
\newacronym{pdu}{PDU}{Packet Data Unit}
\newacronym{pf}{PF}{Proportional Fair}
\newacronym{pgw}{PGW}{Packet Gateway}
\newacronym{phy}{PHY}{Physical}
\newacronym{pbch}{PBCH}{Physical Broadcast Channel}
\newacronym[plural=\gls{mme}s,firstplural=Mobility Management Entities (MMEs)]{mme}{MME}{Mobility Management Entity}
\newacronym{prb}{PRB}{Physical Resource Block}
\newacronym{pss}{PSS}{Primary Synchronization Signal}
\newacronym{pucch}{PUCCH}{Physical Uplink Control Channel}
\newacronym{pusch}{PUSCH}{Physical Uplink Shared Channel}
\newacronym{rach}{RACH}{Random Access Channel}
\newacronym{ran}{RAN}{Radio Access Network}
\newacronym{red}{RED}{Random Early Detection}
\newacronym{rf}{RF}{Radio Frequency}
\newacronym{rlc}{RLC}{Radio Link Control}
\newacronym{rlf}{RLF}{Radio Link Failure}
\newacronym{rrc}{RRC}{Radio Resource Control}
\newacronym{rrm}{RRM}{Radio Resource Management}
\newacronym{rr}{RR}{Round Robin}
\newacronym{rs}{RS}{Remote Server}
\newacronym{rsrp}{RSRP}{Reference Signal Received Power}
\newacronym{rss}{RSS}{Received Signal Strength}
\newacronym{rtt}{RTT}{Round Trip Time}
\newacronym{rw}{RW}{Receive Window}
\newacronym{rx}{RX}{Receiver}
\newacronym{sa}{SA}{standalone}
\newacronym{sack}{SACK}{Selective Acknowledgment}
\newacronym{sap}{SAP}{Service Access Point}
\newacronym{sch}{SCH}{Secondary Cell Handover}
\newacronym{scoot}{SCOOT}{Split Cycle Offset Optimization Technique}
\newacronym{sdma}{SDMA}{Spatial Division Multiple Access}
\newacronym{sinr}{SINR}{Signal to Interference plus Noise Ratio}
\newacronym{sm}{SM}{Saturation Mode}
\newacronym{snr}{SNR}{Signal-to-Noise Ratio}
\newacronym{son}{SON}{Self-Organizing Network}
\newacronym{ss}{SS}{Synchronization Signal}
\newacronym{srs}{SRS}{Sounding Reference Signal}
\newacronym{sss}{SSS}{Secondary Synchronization Signal}
\newacronym{tb}{TB}{Transport Block}
\newacronym{tcp}{TCP}{Transmission Control Protocol}
\newacronym{tdd}{TDD}{Time Division Duplexing}
\newacronym{tdma}{TDMA}{Time Division Multiple Access}
\newacronym{tfl}{TfL}{Transport for London}
\newacronym{tm}{TM}{Transparent Mode}
\newacronym{prr}{PRR}{Packet Reception Ratio}
\newacronym{trp}{TRP}{Transmitter Receiver Pair}
\newacronym{tti}{TTI}{Transmission Time Interval}
\newacronym{ttt}{TTT}{Time-to-Trigger}
\newacronym{tx}{TX}{Transmitter}
\newacronym{ue}{UE}{User Equipment}
\newacronym{ul}{UL}{uplink}
\newacronym{uml}{UML}{Unified Modeling Language}
\newacronym{um}{UM}{Unacknowledged Mode}
\newacronym{utc}{UTC}{Urban Traffic Control}
\newacronym{vm}{VM}{Virtual Machine}
\newacronym{rsrq}{RSRQ}{Reference Signal Received Quality}
\newacronym{rssi}{RSSI}{Received Signal Strength Indicator}
\newacronym{crs}{CRS}{Cell Reference Signal}
\newacronym{v2v}{V2V}{Vehicle-to-Vehicle}
\newacronym{v2i}{V2I}{Vehicle-to-Infrastructure}
\newacronym{v2n}{V2N}{Vehicle-to-Network}
\newacronym{v2x}{V2X}{Vehicle-to-Everything}
\newacronym{vn}{VN}{Vehicular Node}
\newacronym{dsrc}{DSRC}{Dedicated Short Range Communication}
\newacronym{ci}{CI}{context information}
\newacronym{voi}{VoI}{value of information}
\newacronym{gps}{GPS}{Global Positioning System}
\newacronym{qos}{QoS}{Quality of Service}
\newacronym{qoe}{QoE}{Quality of Experience}
\newacronym{ml}{ML}{Machine Learning}
\newacronym{ahp}{AHP}{Analytic Hierarchy Process}
\newacronym{lidar}{LIDAR}{Light Detection and Ranging}
\newacronym{sumo}{SUMO}{Simulation of Urban MObility}
\newacronym{wave}{WAVE}{Wireless Access in Vehicular Environment}
\newacronym{c-its}{C-ITS}{Connected Intelligent Transportation System}
\newacronym{dash}{DASH}{Dynamic Adaptive Streaming over HTTP}
\newacronym{http}{HTTP}{HyperText Transfer Protocol}
\newacronym{nt}{NT}{Non-Terrestrial}
\newacronym{ntc}{NTC}{non-terrestrial communication}
\newacronym{ntn}{NTN}{Non-Terrestrial Network}
\newacronym{hap}{HAP}{High Altitude Platform}
\newacronym{hts}{HTS}{High-Throughput Satellite}
\newacronym{ehts}{EHTS}{Extremely High-Throughput Satellite}
\newacronym{leo}{LEO}{Low Earth Orbit}
\newacronym{meo}{MEO}{Medium Earth Orbit}
\newacronym{geo}{GEO}{Geostationary Earth Orbit}
\newacronym{uav}{UAV}{Unmanned Aerial Vehicle}
\newacronym{nsat}{nSAT}{Nanosatellite}
\newacronym{ehf}{EHF}{extremely high-frequency}
\newacronym{ioe}{IoE}{Internet of Everyone}
\newacronym{gan}{GaN}{Gallium Nitride}
\newacronym{tle}{TLE}{two-line element}
\newacronym{ecdf}{ECDF}{Empirical Cumulative Distribution Function}
\newacronym{fifo}{FIFO}{First-Input First-Output}
\newacronym{isl}{ISL}{Inter-Satellite Link}

\definecolor{steelblue}{RGB}{176,196,222}

\usepackage{hyperref}
\usepackage[capitalize]{cleveref}
\crefname{section}{Sec.}{Secs.}
\usetikzlibrary{decorations}
\usepackage{flushend}

\begin{document}

\title{A Comparative Simulation Study of Satellite Networks: The Case of IRIS$^2$ and Starlink}
\author{Alexander Bonora,~\IEEEmembership{Student Member, IEEE}, \\ Marco~Giordani,~\IEEEmembership{Senior Member, IEEE}, Michele Zorzi,~\IEEEmembership{Fellow, IEEE}
        % <-this % stops a space
%\thanks{Manuscript received March XX, 2024.}
\thanks{The authors are with the Department of Information Engineering, University of Padova. Padova, Italy. (E-mail: \{alexander.bonora, marco.giordani, michele.zorzi\}@dei.unipd.it).\\
This work was partially supported by the European Commission through the European Union’s Horizon Europe Research and Innovation Programme under the Marie Skłodowska-Curie-SE, Grant Agreement No. 101129618, UNITE.}% <-this % stops a space}
}
\maketitle
%TC:ignore
\begin{abstract}
The evolution of \gls{6g} networks relies on satellite-based \glspl{ntn} to extend broadband connectivity to remote and unserved regions, and to support public safety. In this paper we compare two representative and conceptually different satellite constellations, namely Starlink and IRIS$^2$. 
While the former is primarily designed to deliver high-capacity broadband services for civil applications, the latter was mostly conceived to provide a secure, resilient, and sovereign infrastructure for government and critical communications.
%Starlink is a commercial  private Internet constellation by SpaceX, based on dense \gls{leo} satellites. It is primarily designed to deliver high-capacity broadband services for civil applications, with performance targets comparable to those of terrestrial networks.
%In contrast, IRIS$^2$ is a planned public initiative to be deployed by the European Union, based on a multi-layer combination of \gls{leo}, \gls{meo}, and \gls{geo} satellites. It is primarily designed to provide a secure, resilient, and sovereign infrastructure for government and critical communications.
After describing the main technical characteristics of Starlink and IRIS$^2$, we run a comprehensive simulation campaign to evaluate the design trade-offs between the two. Specifically, we evaluate the achievable capacity, the impact of satellite mobility and handover, and identify the capability of each architecture to support global and reliable connectivity.
We also provide design suggestions for possible future IRIS$^2$ deployment extensions.
\end{abstract}
%TC:endignore
%\begin{IEEEkeywords}
%Non-Terrestrial Networks (NTNs); satellite constellations; Starlink; IRIS$^2$; multi-orbit satellite systems.
%\end{IEEEkeywords}

\glsresetall
%TC:ignore
\section{Introduction}
%TC:endignore
\begin{comment}
Future \gls{6g} networks will be designed to enable truly global broadband connectivity. Terrestrial infrastructure remains the backbone of today's Internet. 
Still, the deployment of ground-based networks in remote and unserved areas
is often economically or technologically infeasible, mainly due to harsh environments, limited availability of power and transport infrastructure, and high deployment costs~\cite{Chaoub20216g}.
Moreover, the increasing frequency of natural disasters 
and the growing geopolitical instability expose the vulnerability of existing terrestrial networks, which are not always prepared to ensure reliable communications during emergencies.
In this context, \glspl{ntn} have emerged as a key enabler of global Internet connectivity, particularly where ground infrastructure is absent or compromised~\cite{giordani2021non}.
\end{comment}

Future \gls{6g} networks will rely on \glspl{ntn}~\cite{giordani2021non} to extend broadband connectivity to remote and unserved areas where terrestrial deployment is economically or technologically impractical, and to maintain communications when ground infrastructure is absent or compromised, especially during emergencies~\cite{Chaoub20216g}.

In this context, \gls{leo} satellite mega-constellations, given their proximity to Earth, may provide low-latency, wide-area, and high-capacity Internet connectivity with comparable performance to 4G/5G cellular operators, particularly in rural areas. Notable examples include Amazon Leo, China's Guowang SatNet and Qianfan G60, and Starlink~\cite{lagunas2024low}. %The latter, with over 7\,000 active satellites and millions of customers as of 2025, represents one of the most mature large-scale implementations of satellite-based Internet service~\cite{lagunas2024low}.
Although these mega-constellations are primarily designed for civil applications, they have also been considered for government and military use, e.g., to support command-and-control operations in emergency or warfare scenarios~\cite{bhosale2025assessing}. 
%For instance, during the war in Ukraine in 2022, Starlink terminals were deployed to restore and sustain resilient Internet connectivity after terrestrial networks were disrupted~\cite{bhosale2025assessing}.
However, this prospect raised strategic and regulatory questions at the government level, particularly within the European Union (EU), concerning whether states should rely on a private satellite infrastructure to guarantee public safety %service continuity in case of emergency, 
and handle secure, sensitive, or sovereignty-critical data.

To promote strategic autonomy in space, the EU has initiated the development of public satellite networks, most notably via the Infrastructure for Resilience, Interconnectivity and Security by Satellite (IRIS$^2$) initiative, with initial services expected around 2030.
Unlike commercial mega-constellations, 
which are primarily designed for mass-market connectivity, 
IRIS$^2$ will deploy a more limited number of satellites (around 300), and therefore is not intended to match the same large-scale capacity and global coverage performance.
Instead, it is meant to provide secure, sovereign, and reliable communications for dedicated sensitive stakeholders, including governmental authorities, defense and military organizations, and critical infrastructure operators~\cite{shankar2024iris2}. Once fully operational, IRIS$^2$ will also provide services to businesses, citizens, and unserved regions.
Notably, IRIS$^2$ will adopt a multi-orbit architecture combining both \gls{leo} satellites, to ensure low latency, and \gls{meo} satellites, to provide large-scale coverage and service continuity. 
%To achieve these goals, IRIS$^2$ will implement sophisticated security features, potentially including Quantum Key Distribution (QKD), and adopt a multi-orbit architecture combining both Low \gls{leo} satellites, to ensure low latency, with \gls{meo} satellites, to provide large-scale coverage and service continuity. 
%This hybrid approach aims to balance resilience, coverage, and the protection of sensitive communications.

\begin{comment}
\begin{figure*}[t]
    \centering
    \includegraphics[width=\textwidth]{Starlink and Iris comparison/figures/scenario_image.png}
    \caption{An illustration of the IRIS$^2$ and Starlink constellation models.}
    \label{fig:scenario}
\end{figure*}
\end{comment}

Given the high estimated cost (approximately 10.5 billion Euros, of which 6.5 billion from public EU funds) and the associated strategic implications, it is essential to rigorously evaluate the performance of IRIS\textsuperscript{2}, also in comparison with other private satellite systems.
In this work, we use Starlink as the reference benchmark, as it is currently the most representative, well-documented, and operationally mature commercial mega-constellation, and has been at the center of the European debate on sovereignty in space that motivated the development of IRIS\textsuperscript{2}.
To this end, in this paper we first describe the main technical
characteristics of Starlink and IRIS\textsuperscript{2} (Sec.~\ref{sub:techincal-description}). Then, we introduce a unified system model %(including orbital mobility, deployment, channel, and capacity models) 
to evaluate both under similar assumptions and conditions (Sec.~\ref{sec:system-model}). Based on this framework, we evaluate via simulations some design trade-offs between Starlink and IRIS\textsuperscript{2} (Sec.~\ref{sec:results}). 
%In particular, we assess the per-cell and per-user achievable throughput, as well as the coverage performance, and evaluate the capability of the two systems to support global, reliable, and broadband Internet connectivity.
Our results indicate that, while Starlink achieves superior connectivity performance in absolute terms, as expected, IRIS\textsuperscript{2} may provide competitive advantages in specific strategic scenarios, contributing to the EU technological autonomy and geopolitical sovereignty.

\section{A Technical Description of \\ Starlink and IRIS$^2$ Constellations}
\label{sub:techincal-description}

\subsection{Starlink for Global Broadband Connectivity}
%TC:endignore
Starlink is a commercial \gls{leo} constellation providing high-speed, low-latency Internet connectivity worldwide.
At the time this study was done, it comprised about 7\,000 satellites, mostly deployed at 540--570 km, though the long-term plan is to have up to 42\,000 to ensure near-global coverage~\cite{10.3389/frcmn.2021.643095}. 
The close proximity to Earth ensures \glspl{rtt} of 25–35 ms, making real-time applications such as video conferencing, cloud access, and gaming feasible.
Satellites operate primarily in the Ku-band (around 12-18 GHz) and Ka-band (around 26.5–40 GHz), with an expected download (upload) Internet speed from around 48 (10) to 265 (26) Mbps according to the Starlink specifications, so with comparable performance  to terrestrial cellular networks.

Starlink employs a proprietary phased-array flat-panel antenna with electronic beam steering, which enables rapid beam switching and seamless satellite handovers. 
The new generation of Starlink satellites (regenerative payload) also integrates Optical Inter-Satellite Links (OISLs).
These links create a global mesh network for data to be relayed entirely through space without returning to ground stations, reducing latency on long routes, and extending coverage to remote regions such as oceans or polar areas where gateway infrastructure cannot be deployed~\cite{9393372}.
Moreover, Starlink supports frequency reuse across different (geographically separated) beams, which is essential to support millions of users with a limited spectrum.

While Starlink is primarily designed for mass-market broadband connectivity, its infrastructure inherently supports dual-use scenarios for government and tactical operations, e.g., for disaster recovery during public emergencies, remote infrastructure monitoring, and defense communications~\cite{bhosale2025assessing}.
However, the opportunity to rely on private satellite operators to support public safety may raise concerns regarding data security, loss of sovereign control, and the potential influence of commercial interests over public priorities, especially during crises.

\subsection{IRIS\textsuperscript{2} for Secure and Resilient Connectivity}
IRIS$^2$ was announced under the EU Secure Connectivity Programme 2023–2027 as a sovereign and resilient satellite infrastructure. It will support secure governmental communications, crisis management, protection of critical infrastructures, surveillance and defense operations, and broadband connectivity to businesses, citizens and unserved~regions.

IRIS$^2$ is conceived as a multi-orbit constellation combining 264 LEO satellites at around 1\,200 km, 10 Low-LEO satellites at around 400 km, 18 MEO satellites between approximately 5\,000 and 10\,000 km, and 10 existing \gls{geo} satellites. Compared to a single dense \gls{leo} shell, such as that of Starlink, this multi-orbit architecture is designed to balance capacity with coverage, resiliency, and centralized control. 
In particular, the \gls{leo} and Low-\gls{leo} layers provide high-speed low-latency connectivity given their proximity to Earth, the \gls{meo} layer supports large-scale continuous and resilient coverage with a substantially smaller number of satellites, and GEO satellites provide a secure, high-capacity backbone for gateway access and network coordination. This hierarchical organization is complemented by a two-layer software-defined architecture, inspired by~\cite{wei2025iris}.

IRIS\textsuperscript{2} satellites will operate in the Ku-band (12--18~GHz), Ka-band (26.5--40~GHz), and Q/V-band (Q: 33--50~GHz, V: 40--75~GHz), and in dedicated secure bands for classified or mission-critical services.
The system will feature regenerative payloads for onboard data processing, electronically steerable multi-beam antennas, digital channelizers for flexible bandwidth allocation, OISL capability for inter-satellite routing, and rapid network reconfiguration.
Notably, the IRIS\textsuperscript{2} infrastructure follows a ``security-by-design'' approach, integrating end-to-end encryption, anti-jamming mechanisms, and robust cryptographic solutions across the space and ground segments. 

At the time of writing, the programme is in the early~implementation stages, with initial satellite launches planned around 2029, and full operational capability targeted by 2030.
Unlike Starlink, IRIS$^2$ was not developed to maximize mass-market capacity. Rather, it was primarily designed to prioritize a comparatively smaller number of high-priority users in selected crisis scenarios. This difference motivates the comparative analysis formalized in this paper.

\section{Unified System Model}
\label{sec:system-model}
%TC:endignore

In this section we present a unified system model that we integrated in a simulation framework to evaluate both Starlink and IRIS$^2$ architectures under similar assumptions and conditions. We describe the deployment scenario (Sec.~\ref{sub:scenario}), and the channel (Sec.~\ref{sub:channel}), orbital (Sec.~\ref{sub:orbital}), and association (Sec.~\ref{sub:association}) models.

%TC:ignore
\subsection{Deployment Scenario}
\label{sub:scenario}
%TC:endignore

 Our scenario considers a constellation of satellites, each serving $n$ \glspl{ue} distributed across different ground regions. Specifically, the Earth is represented as a sphere, and satellites are modeled as point objects defined by their geodetic coordinates (latitude, longitude, and altitude).  We analyze and compare the two architectures described in Sec.~\ref{sub:techincal-description}.
For Starlink, the constellation model consists of $L^{\rm S}$ LEO satellites at altitude $h^{\rm S}_L$. Each satellite shapes cells of radius $R^{\rm S}_L$, and is equipped with antennas forming up to $N^{\rm S}_L$ high-capacity spot beams on the ground, where the effective coverage area of each beam partially overlaps with neighboring beams and, in some cases, with beams from other satellites in similar orbital shells. Consequently, a \gls{ue} may be simultaneously visible to multiple satellites at different elevation angles. Our conservative assumption is that the coverage radius of each spot beam cell is modeled as one-tenth of the cell radius, i.e., $R^{\rm S}_L/10$.
For IRIS$^2$, the constellation consists of $L^{\rm I}$ LEO and Low-LEO satellites and  $M^{\rm I}$ MEO satellites at altitudes $h^{\rm I}_L$ and $h^{\rm I}_M$, shaping $N^{\rm I}_L$ and $N^{\rm I}_M$ high-capacity spot beams on the ground, forming cells of radius $R^{\rm I}_L/10$ and $R^{\rm I}_M/10$, respectively.

\subsection{Channel Model}
\label{sub:channel}
%TC:endignore

The ground-to-satellite channel is modeled based on the 3GPP 5G NR-NTN specifications in~\cite{38821,38811}. %It is primarily dominated by a \gls{los} component.
Large-scale effects include free-space path loss %which depends on the carrier frequency and the ground-to-satellite distance, 
and additional attenuation factors such as atmospheric absorption,
%(including gaseous absorption from oxygen and water vapor), 
rain fading, shadowing, and scintillation. These effects depend on the propagation environment and the elevation angle, i.e., the angle between the satellite and the ground \gls{ue}, with lower elevation angles generally resulting in higher attenuation and increased blockage probability. 
%The model also accounts for scintillation effects due to irregularities in the ionosphere and the troposphere, which introduce rapid fluctuations in the amplitude and phase of the received signal. These effects are more pronounced at lower elevation angles and higher frequency.
Given the limited scattering environment, small-scale multi-path fading is generally less severe than in terrestrial networks, except in urban scenarios. The temporal characteristics of the channel are primarily affected by the satellite dynamics, especially for \gls{leo} satellites, leading to significant Doppler shifts affecting both frequency synchronization and channel estimation performance.

Satellite antennas are modeled as circular aperture antennas, while \glspl{ue} employ \gls{upa} or Very Small Aperture Terminal (VSAT) antennas (representing a Starlink-like phased-array flat-panel antenna), as specified in~\cite{38821}.

The channel quality is measured in terms of \gls{snr}, which is used to obtain the theoretical capacity based on Shannon's formula.
For cells where multiple satellites are simultaneously visible, the aggregate cell capacity is measured as the sum of the capacities from all visible satellites in the cell. Notably, consistent with the 3GPP NR-NTN specifications~\cite{38821,38811}, overlapping beams are allocated non-overlapping frequency resources, which prevents mutual interference.
The per-UE capacity is the capacity of a single satellite beam, divided by the number of UEs in the~beam.
%TC:ignore
%For cells where multiple satellites are simultaneously visible, the aggregate capacity is measured as the sum of the capacities from all visible satellites in the cell. The effective per-UE capacity is the cell capacity divided by the number of UEs in the cell.
%This formulation models a homogeneous allocation of resources, allowing for straightforward comparison of load balancing and beam overlap effects in both Starlink and IRIS\textsuperscript{2} constellations.

%Note that the aggregate capacity does not imply that a \gls{ue} combines signals from multiple satellites: each \gls{ue} is served by exactly one satellite and beam (Sec.~\ref{sub:association}), and, consistent with the 3GPP NR-NTN framework~\cite{38821,38811}, overlapping beams are allocated non-overlapping frequency resources, so the summed capacities represent additional, non-interfering resources rather than mutually interfering signals.

%TC:ignore
\begin{table*}[t!]
\centering
\caption{Simulation parameters.}
\label{tab:params}
\begin{tabular}{@{}lll@{}}
\toprule
{Symbol} & {Definition} & {Value} \\
\midrule
EIRP$_{\text{UL}}$ & UE isotropically radiated power & \SI{45.01}{dBW} (\gls{leo}); \SI{48.01}{dBW} (\gls{meo}) \\
$(G/T)_{\text{UL}}$ & Satellite gain-to-noise-temperature ratio & \SI{5.0}{dB/K} (\gls{leo}); \SI{8.0}{dB/K} (\gls{meo}) \\
\midrule
EIRP$_{\text{DL}}$ & Satellite isotropically radiated power & \SI{36.02}{dBW} \\
$(G/T)_{\text{DL}}$ & UE gain-to-noise-temperature ratio & \SI{21.44}{dB/K} \\
\midrule
$L^{\rm S}$ & Number of Starlink \gls{leo} & 6937 \\
$h^{\rm S}_L$ & Average altitude for Starlink \gls{leo} & {600} {km} \\
$R^{\rm S}_L$ & Average cell radius for Starlink \gls{leo} & {600} {km} \\
\midrule
$L^{\rm I}$ & Number of IRIS$^2$ \gls{leo} & 264 (LEO) + 10 (Low-LEO) \\
$M^{\rm I}$ & Number of IRIS$^2$ \gls{meo} & 18 \\
$h^{\rm I}_L$ & Altitude for IRIS$^2$ \gls{leo} & [400,1\,400] km \\
$h^{\rm I}_M$ & Altitude for IRIS$^2$ \gls{meo} & [7\,500,8\,500] km \\
$R^{\rm I}_L$ & Cell radius for IRIS$^2$ \gls{leo} & [600,1\,200] km \\
$R^{\rm I}_M$ & Cell radius for IRIS$^2$ \gls{meo} & 2\,500 km \\
$N^{\rm I}_L;N^{\rm I}_M;N^{\rm S}_L$ & Total number of spot beams per satellite cell & 61 \\
\midrule
$n_{L}$ & Average number of IRIS$^2$ \gls{leo} satellites per \gls{meo} & 28.5 \\
$\tau_p$ & Handover processing time & 0--30 ms \\
$\alpha$ & Logical handover downscaling factor & 0.3\\
\midrule
$B$ & Bandwidth per spot beam & \SI{400}{MHz} \\
$R_E$ & Earth's radius & \SI{6371}{km} \\
\bottomrule
\end{tabular}
\end{table*}

\subsection{Orbital Model}
\label{sub:orbital}
%In this section we describe how the position and dynamics of the satellites in the Starlink and {IRIS$^2$} constellations are determined.
%Specifically, the satellite positions are transformed into the Earth-Centered, Earth-Fixed (ECEF) reference frame. This representation is used to determine the distance between the \glspl{ue} and the satellites, as well as the corresponding \gls{los} condition, which are required for evaluation of the \gls{snr}~\cite{10978515}.
%TC:endignore

\paragraph{Starlink}
The orbital configuration of the Starlink constellation is designed to ensure continuous global coverage. Satellites operate in a circular \gls{leo}, and are distributed across multiple orbital shells and inclined planes. This configuration allows multiple satellites to be visible from most locations on Earth, ensuring redundancy and reducing service interruptions. 
Starlink's orbital parameters are derived in real time from publicly available \gls{tle} data, which describe the satellites' trajectories relative to the Earth, and are used to calculate the instantaneous position, velocity, and orbital phase of each satellite in the~constellation.

\paragraph{IRIS$^2$}
Since the IRIS$^2$ constellation is still under development, direct \gls{tle} data are not yet available.  Therefore, orbital parameters are generated based on a Walker–$\Delta$ constellation model,
%TC:ignore
%ensuring a uniform global satellite distribution comparable to that of Starlink~\cite{engproc2025088057}. Satellites positions are first derived analytically, and subsequently converted into latitude, longitude, and altitude coordinates.
%TC:endignore
which specifies how satellites are distributed across multiple orbital planes around the Earth~\cite{engproc2025088057}. 

The model is commonly referred to as Walker(${N}$:${P}$:${F}$), where $N$ is the total number of satellites to be deployed, $P$ is the number of orbital planes, and $F$ is a phasing parameter controlling the relative angular position of satellites between adjacent planes. 
First, the Right Ascension of the Ascending Node (RAAN) determines the orientation of each orbital plane in space. Specifically, the RAAN is defined as the angle in the equatorial plane between a reference direction (typically the vernal equinox) and the ascending node, i.e., the point where a satellite crosses the equatorial plane from south to north.
The orbital planes are assumed to be uniformly distributed over the interval $[0,2\pi)$ in terms of RAAN, so that consecutive planes have an angular separation of $2\pi/P$. This configuration provides an approximately uniform longitudinal distribution of orbital planes around the Earth.
Once the orbital planes are defined, $N/P$ satellites are uniformly distributed along each orbit. Then, the phasing parameter $F$ introduces a relative angular offset between satellites in adjacent orbital planes, resulting in a more uniform spatial distribution of satellites across the constellation, comparable to that of Starlink.
 Finally, satellite positions are subsequently converted into latitude, longitude, and altitude coordinates.

%TC:ignore
%In general, the spatial relationship between a user terminal and a satellite is defined by three main quantities: the slant range, the central angle, and the elevation angle. Higher elevation angles correspond to shorter slant ranges and stronger link conditions, while lower elevation angles typically result in greater atmospheric attenuation and increased path loss. 

\begin{comment}
Let $N_s = \lceil N / P \rceil$ denote the number of satellites per orbital plane. For the $k$-th orbital plane and the $j$-th satellite ($j = 0, 1, \dots, N_s - 1$), the along-track phase or mean anomaly is defined as
\begin{align}
\omega_{k,j} 
&= \left( 
\frac{2\pi j}{N_s} 
+ \frac{2\pi F k}{N}
\right) \bmod 2\pi,
\label{eq:walker_phase}
\end{align}
where $F$ introduces a relative phase shift between successive planes, thereby generating the well-known \emph{Walker--$\delta$ pattern}.  
This arrangement produces an approximately uniform global coverage pattern, as satellites are evenly spaced within each plane and the planes themselves are uniformly distributed in RAAN.
\end{comment}

\subsection{Satellite Association Model}
\label{sub:association}
%The simulation framework generates a three-dimensional spatial representation of the Earth and the satellite constellation, enabling precise geometric modeling of \glspl{ue}, satellite positions, and beam footprints. For each \gls{ue}, the set of visible satellites is dynamically determined based on \gls{los} constraints and the Earth’s curvature. Subsequently, the elevation angle between each \gls{ue} and its visible satellites is computed to assess link quality and coverage probability.

 %By associating each \gls{ue} with the most suitable satellite and beam based on elevation angle and signal conditions, we estimate the available throughput for individual \gls{ue}s and for clusters of \gls{ue}s distributed across different geographic regions.

 %The user’s position on the ground, defined by its latitude and longitude, determines the set of satellites in view at any given time, depending on elevation constraints and visibility windows.

 %Specifically, the orbital planes are designed to ensure that, at any location, at least one satellite is always visible above a given minimum elevation threshold.

 %The simulation framework generates a three-dimensional spatial representation of the Earth and the satellite constellation, enabling precise geometric modeling of \glspl{ue}, satellite positions, and beam footprints.
 %TC:endignore
 A satellite is in visibility with respect to a certain ground \gls{ue} if it is in \gls{los}, accounting for the Earth’s curvature, and the elevation angle between the two is above a minimum predefined threshold.
 %TC:ignore
 %For each \gls{ue}, the set of visible satellites is dynamically determined by considering \gls{los} constraints and the Earth’s curvature. 
 %Based on this visibility set, the elevation angle between each \gls{ue} and the candidate satellites is computed, serving as a key metric to evaluate link quality and coverage probability.
%The geographical position of each \gls{ue}, defined by its latitude and longitude, directly determines the subset of satellites within its visibility region at any given time, subject to minimum elevation angle constraints.
%TC:endignore
The constellation is designed to ensure continuous service availability, such that at least one satellite is always visible from any ground location at all times.
%TC:ignore
%In this work, we assume that the access link is not a limiting factor, i.e., it is capable of handling the offered traffic load without congestion. Furthermore, handover procedures are assumed to be seamless, with negligible delay or signaling overhead. Under these assumptions, satellites are considered to have full knowledge of the network state, including the set of currently connected \glspl{ue} and their respective link conditions.
%TC:endignore
Two association models are considered. In the first model (``Best SNR''), each \gls{ue} is associated with the satellite in visibility (and the corresponding beam) that provides the best signal quality, measured in terms of the \gls{snr}. 
In the second model (``Load Balancing''), each \gls{ue} is associated with the satellite in visibility (and the corresponding beam) that maximizes its per-UE capacity, therefore taking into account both channel conditions and the availability of resources, that is the number of UEs in the cell.
This approach assumes that the satellites have full knowledge of the network state, including the set of connected \glspl{ue} and their link conditions.

%TC:ignore
\section{Simulation Results}
\label{sec:results}
\subsection{Simulation Parameters}
The simulation parameters are reported in Table~\ref{tab:params}.
%TC:endignore
For Starlink, we consider $L^{\rm S}=6\,937$ LEO satellites at altitude $h^{\rm S}_L=600$ km, shaping cells of radius $R^{\rm S}_L=600$ km.
For IRIS$^2$, we consider $L^{\rm I}=10+264=274$ Low-LEO and LEO satellites at altitudes $h^{\rm I}_L\in[400,1\,400]$ km, shaping cells of radius $R^{\rm I}_L\in[600,1\,200]$ km, and $M^{\rm I}=18$ MEO satellites at altitudes $h^{\rm I}_M\in[7\,500,8\,500]$ km, shaping cells of radius $R^{\rm I}_M=2\,500$ km.
Satellites form $N^{\rm I}_L=N^{\rm I}_M=N^{\rm S}_L=61$ high-capacity partially-overlapping spot beams on the ground, covering around one-tenth of the cell radius. Each spot beam is allocated a bandwidth of 400 MHz. 

In uplink, the isotropically radiated power (EIRP, which accounts for the antenna transmit power, the cable loss, and the transmit antenna gain) of the UEs is 45.01 and 48.01 dBW for transmissions toward \gls{leo} and \gls{meo} satellites, respectively, while the receive antenna-gain-to-noise-temperature ($G/T$) at the LEO and MEO satellites is 5.0 and 8.0 dB/K, respectively. 
In downlink, the EIRP of the satellites is 36.02 dBW, while for UEs we set $G/T=21.44$~dB/K. 

Satellites move according to the orbital model described in Sec.~\ref{sub:orbital}, with velocities of 7.56 km/s and 3.07 km/s for \gls{leo} and \gls{meo} satellites, respectively, while \glspl{ue} are stationary. 

%TC:ignore
\subsection{Numerical Results}

\begin{figure}[t!]
\centering
\begin{tikzpicture}

\definecolor{darkgray176}{RGB}{176,176,176}
\definecolor{darkorange25512714}{RGB}{255,127,14}
\definecolor{lightgray204}{RGB}{204,204,204}
\definecolor{steelblue31119180}{RGB}{31,119,180}

\begin{axis}[
  clip=false,
  tick pos=left,
  x grid style={darkgray176},
  tick label style={font=\footnotesize},
  ylabel style={font=\footnotesize},
  xlabel style={font=\footnotesize},
  xmin=-0.5, xmax=3.5,
  xtick style={color=black},
  xtick={0,1,2,3},
  xticklabels={2\,500,5\,000,10\,000,30\,000},
  y grid style={darkgray176},
  ytick style={color=black},
  legend cell align={left},
  legend columns=2,
  % legenda sopra l'asse: con 4 voci (2 metriche x 2 sistemi) una legenda
  % interna si sovrapporrebbe alle barre aggregate di Starlink (61.6 su 80)
  legend style={font=\footnotesize,fill opacity=0.8, draw opacity=1, text opacity=1, draw=lightgray204,
                at={(0.5,1.03)}, anchor=south},
  xlabel={Number of UEs ($n$)},
  ylabel=\textcolor{steelblue31119180}{Aggregate cell capacity [Gbps]},
  ymajorgrids,
  ymin=0, ymax=80
]
\draw[draw=black,fill=steelblue31119180] (axis cs:-0.4,0) rectangle (axis cs:-0.2,12.5078873936263);
\draw[draw=black,fill=steelblue31119180] (axis cs:0.6,0) rectangle (axis cs:0.8,12.5078873936263);
\draw[draw=black,fill=steelblue31119180] (axis cs:1.6,0) rectangle (axis cs:1.8,12.5078873936263);
\draw[draw=black,fill=steelblue31119180] (axis cs:2.6,0) rectangle (axis cs:2.8,12.5078873936263);
\draw[draw=black,fill=steelblue31119180,postaction={pattern=north west lines}] (axis cs:-0.2,0) rectangle (axis cs:0,61.577553655258);
\draw[draw=black,fill=steelblue31119180,postaction={pattern=north west lines}] (axis cs:0.8,0) rectangle (axis cs:1,61.577553655258);
\draw[draw=black,fill=steelblue31119180,postaction={pattern=north west lines}] (axis cs:1.8,0) rectangle (axis cs:2,61.577553655258);
\draw[draw=black,fill=steelblue31119180,postaction={pattern=north west lines}] (axis cs:2.8,0) rectangle (axis cs:3,61.577553655258);
\addlegendimage{ybar,area legend,draw=black,fill=steelblue31119180}
\addlegendentry{Aggregate cell capacity}
\addlegendimage{ybar,area legend,draw=black,fill=darkorange25512714}
\addlegendentry{Per-UE capacity}
\addlegendimage{ybar,area legend,draw=black,fill=white}
\addlegendentry{IRIS$^2$}
\addlegendimage{ybar,area legend,draw=black,fill=white,postaction={pattern=north west lines}}
\addlegendentry{Starlink}
\end{axis}

\begin{axis}[
  axis y line=right,
  tick align=outside,
  x grid style={darkgray176},
  xmin=-0.5, xmax=3.5,
  xtick pos=left,
  xtick style={color=black},
  xtick=\empty,
  xticklabels=\empty,
  y grid style={darkgray176},
  tick label style={font=\footnotesize},
  ylabel=\textcolor{darkorange25512714}{Per-UE capacity [Mbps]},
  ylabel style={at={(axis description cs:1.1,0.5)},font=\footnotesize},
  ymin=0, ymax=120,
  ytick pos=right,
  ytick style={color=black},
  yticklabel style={anchor=west}
]
\draw[draw=black,fill=darkorange25512714] (axis cs:0,0) rectangle (axis cs:0.2,80.4124627412398);
\draw[draw=black,fill=darkorange25512714] (axis cs:1,0) rectangle (axis cs:1.2,40.2062313706199);
\draw[draw=black,fill=darkorange25512714] (axis cs:2,0) rectangle (axis cs:2.2,20.10311568531);
\draw[draw=black,fill=darkorange25512714] (axis cs:3,0) rectangle (axis cs:3.2,6.70103856176999);
\draw[draw=black,fill=darkorange25512714,postaction={pattern=north west lines}] (axis cs:0.2,0) rectangle (axis cs:0.4,110.819709380875);
\draw[draw=black,fill=darkorange25512714,postaction={pattern=north west lines}] (axis cs:1.2,0) rectangle (axis cs:1.4,55.4098546904374);
\draw[draw=black,fill=darkorange25512714,postaction={pattern=north west lines}] (axis cs:2.2,0) rectangle (axis cs:2.4,27.7049273452187);
\draw[draw=black,fill=darkorange25512714,postaction={pattern=north west lines}] (axis cs:3.2,0) rectangle (axis cs:3.4,9.23497578173957);
\end{axis}

\end{tikzpicture}
\vspace{-0.33cm}
\caption{Downlink aggregate (left axis) and per-UE (right axis) capacity for IRIS\textsuperscript{2} vs. Starlink, as a function of the number of UEs. UEs are associated with the satellite that provides the highest SNR (``Best SNR'').}
\vspace{-0.5cm}
\label{fig:ue_aggr}
\end{figure}
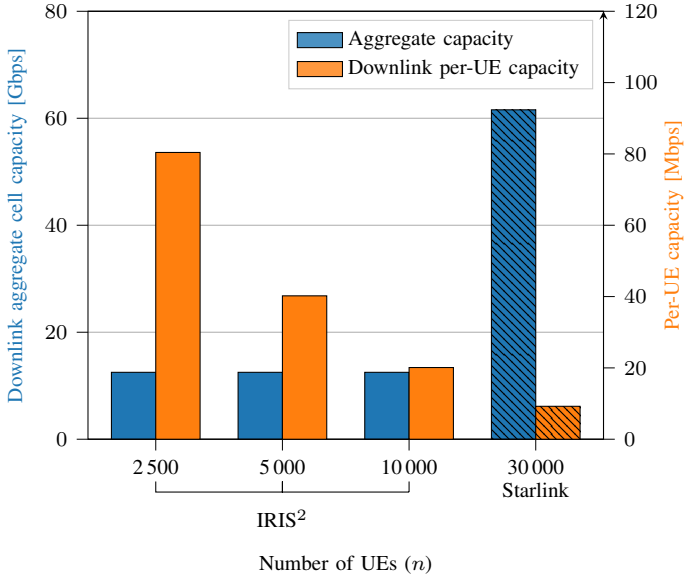

%TC:endignore
\paragraph{Capacity}
In Fig.~\ref{fig:ue_aggr} we plot the downlink aggregate cell capacity (left axis), computed as the sum of the capacities from all visible satellites in a cell.
We observe that Starlink achieves approximately 4.5$\times$ higher aggregate capacity than IRIS$^2$.
This is expected given the density of the Starlink constellation, which comprises more than 23$\times$ the number of satellites than IRIS$^2$. The resulting effect is twofold. First, each cell is, on average, served by more simultaneously visible satellites, which provide additional transmission resources. Second, the average distance between the UEs and the serving satellites decreases when the constellation is more dense, translating into higher received power (and SNR) and capacity.

In Fig.~\ref{fig:ue_aggr} we also plot the downlink per-\gls{ue} capacity (right axis) as a function of the number of \glspl{ue} $n\in\{2\,500,5\,000,10\,000,30\,000\}$ in the cell, for both IRIS$^2$ and Starlink. \Gls{ue}s are associated with the satellite that provides the highest SNR (``Best SNR'' in Sec.~\ref{sub:association}). 
First, we observe that the per-\gls{ue} capacity decreases as $n$ increases for both systems.
Second, Starlink achieves higher per-\gls{ue} capacity than IRIS$^2$, given the additional available transmission resources. However, the gap between the two systems is only around $1.38\times$ for all values of $n$, vs. around $4.5\times$ in terms of aggregate capacity.
Finally, notice that Starlink is a commercial broadband system, designed to provide broadband Internet connectivity at scale to a large population of \gls{ue}s per cell, with $n$ up to tens of thousands. On the contrary, IRIS\textsuperscript{2} is primarily conceived as a sovereign and secure communication infrastructure, supporting connectivity services for a comparatively limited number of institutional, governmental and mission-critical users.
Under a more realistic operating regime, with $n$ ranging from approximately $2\,500$ to $10\,000$, the per-\gls{ue} capacity of IRIS\textsuperscript{2} is around 20-60~Mbps, which is comparable to the connectivity performance of current terrestrial 4G/5G~networks.

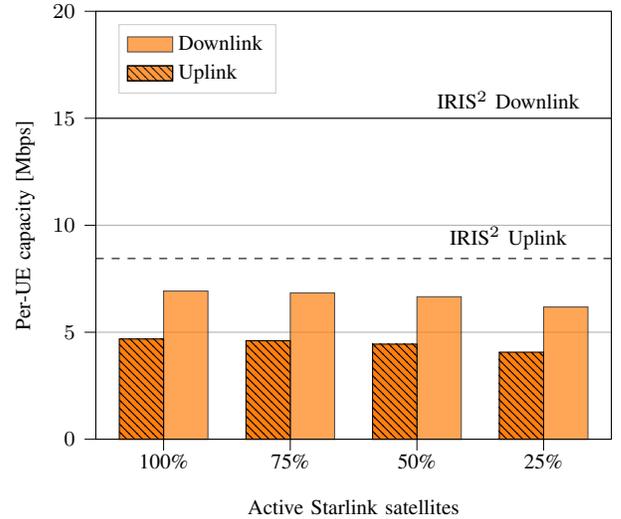
\begin{figure}[t!]
\centering
% This file was created with tikzplotlib v0.9.17.
\begin{tikzpicture}
\definecolor{darkgray176}{RGB}{176,176,176}
\definecolor{darkorange25512714}{RGB}{255,127,14}
\definecolor{lightgray204}{RGB}{204,204,204}
\definecolor{steelblue31119180}{RGB}{31,119,180}

\begin{axis}[
clip=false,
legend cell align={left},
legend style={font=\footnotesize,fill opacity=0.8, draw opacity=1, text opacity=1, draw=lightgray204,at={(0.35,0.97)}},
tick align=outside,
tick pos=left,
x grid style={darkgray176},
xmin=-0.535, xmax=3.535,
xtick style={color=black},
xtick={0,1,2,3},
xticklabels={100\%,75\%,50\%,25\%},
xlabel={\parbox{4cm}{\centering Active Starlink satellites}},
tick label style={font=\footnotesize},
ylabel style={font=\footnotesize},
xlabel style={font=\footnotesize},
xticklabel style={yshift=0.5ex},
xlabel style={yshift=-0.5em},
y grid style={darkgray176},
ylabel={Per-UE capacity [Mbps]},
ymajorgrids,
ymin=0, ymax=20,
ytick style={color=black}
]

\draw[black]
    (axis cs:-0.535,15) -- (axis cs:3.535,15)
    node[pos=0.8, anchor=south,font=\footnotesize] {IRIS$^2$ Downlink};

\draw[dashed, black]
    (axis cs:-0.535,8.4442761575218) -- (axis cs:3.535,8.4442761575218)
    node[pos=0.8, anchor=south,font=\footnotesize] {IRIS$^2$ Uplink};

%\draw[draw=black,fill=steelblue31119180,opacity=0.7] (axis cs:-0.35,0) rectangle (axis cs:0,15.077336763984);
%\addlegendimage{ybar,area legend,draw=black,fill=steelblue31119180,opacity=0.7}
%\addlegendentry{Iris² Downlink}

%\draw[draw=black,fill=steelblue31119180,opacity=0.7] (axis cs:0.65,0) rectangle (axis cs:1,15.077336763984);
%\draw[draw=black,fill=steelblue31119180,opacity=0.7] (axis cs:1.65,0) rectangle (axis cs:2,15.077336763984);
%\draw[draw=black,fill=steelblue31119180,opacity=0.7] (axis cs:2.65,0) rectangle (axis cs:3,15.077336763984);
\draw[draw=black,fill=darkorange25512714,opacity=0.7] (axis cs:2.77555756156289e-17,0) rectangle (axis cs:0.35,6.92623183631414);
\addlegendimage{ybar,area legend,draw=black,fill=darkorange25512714,opacity=0.7}
\addlegendentry{Downlink}

\draw[draw=black,fill=darkorange25512714,opacity=0.7] (axis cs:1,0) rectangle (axis cs:1.35,6.83957020605298);
\draw[draw=black,fill=darkorange25512714,opacity=0.7] (axis cs:2,0) rectangle (axis cs:2.35,6.65853694815928);
\draw[draw=black,fill=darkorange25512714,opacity=0.7] (axis cs:3,0) rectangle (axis cs:3.35,6.18058636652988);
%\draw[draw=black,fill=steelblue31119180,postaction={pattern=north west lines}] (axis cs:-0.2975,0) rectangle (axis cs:-0.0525,8.4442761575218);
%\addlegendimage{ybar,area legend,draw=black,fill=steelblue31119180,postaction={pattern=north west lines}}
%\addlegendentry{Iris² Uplink}

%\draw[draw=black,fill=steelblue31119180,postaction={pattern=north west lines}] (axis cs:0.7025,0) rectangle (axis cs:0.9475,8.4442761575218);
%\draw[draw=black,fill=steelblue31119180,postaction={pattern=north west lines}] (axis cs:1.7025,0) rectangle (axis cs:1.9475,8.4442761575218);
%\draw[draw=black,fill=steelblue31119180,postaction={pattern=north west lines}] (axis cs:2.7025,0) rectangle (axis cs:2.9475,8.4442761575218);
\draw[draw=black,fill=darkorange25512714,postaction={pattern=north west lines}] (axis cs:-0.35,0) rectangle (axis cs:0,4.68588920902091);
\addlegendimage{ybar,area legend,draw=black,fill=darkorange25512714,postaction={pattern=north west lines}}
\addlegendentry{Uplink}

\draw[draw=black,fill=darkorange25512714,postaction={pattern=north west lines}] (axis cs:0.65,0) rectangle (axis cs:1,4.602737765997);
\draw[draw=black,fill=darkorange25512714,postaction={pattern=north west lines}] (axis cs:1.65,0) rectangle (axis cs:2,4.44539625579657);
\draw[draw=black,fill=darkorange25512714,postaction={pattern=north west lines}] (axis cs:2.65,0) rectangle (axis cs:3,4.06911672585122);
\end{axis}

\end{tikzpicture}
\caption{Per-\gls{ue} downlink and uplink capacity for IRIS$^2$ vs. Starlink, as a function of the percentage of active Starlink satellites. IRIS$^2$ serves $n=10\,000$ mission-critical \glspl{ue}, while Starlink serves $n=30\,000$ mass-market \glspl{ue}. \gls{ue}s are associated with the satellite that provides the highest SNR.}
\vspace{-0.5cm}
\label{fig:percent_bs}
\end{figure}

\paragraph{Robustness}
Another advantage of IRIS$^2$ is that, unlike private commercial systems like Starlink, it is governed and controlled by the EU, ensuring sovereign autonomy in line with public interests and objectives.
We consider an emergency scenario where a percentage of Starlink satellites becomes unavailable, either due to external disruptions (e.g., jamming or cyberattacks) or deliberate service restrictions imposed by the operator in sensitive contexts~\cite{abels2024private}. 
%This scenario is realistic. For example, in February 2023, SpaceX's COO Gwynne Shotwell announced that Starlink service was partially restricted for certain Ukrainian military drone operations, arguing that the infrastructure was \emph{``never intended to be weaponized''} or to be used \emph{''for offensive purposes''}~\cite{abels2024private}.
In Fig.~\ref{fig:percent_bs}, we plot the per-UE capacity in both uplink and downlink as a function of the percentage of active Starlink satellites. We set $n=30\,000$ for Starlink, and $n=10\,000$ for IRIS$^2$.
As this percentage decreases, the per-UE capacity also decreases (up to $-15\%$ in uplink)  since UEs are, in general, forced to connect to more distant satellites, which degrades the quality (SNR) of the channel and the link capacity.
This result highlights the sensitivity of large-scale commercial constellations to operational constraints, and motivates the interest to develop complementary public satellite infrastructures like IRIS\textsuperscript{2}.

%TC:ignore
%TC:endignore
Additional offline simulations over the European landmass (not reported here for lack of space) show that Starlink provides a higher average number of visible satellites than IRIS$^2$ (10.41 vs.~7.67). However, IRIS$^2$ exhibits a more uniform spatial distribution across the European territory. For example, in Northern Scandinavia, Starlink provides an average of only 0.46 visible satellites, vs. 3.14 for IRIS$^2$. 
%Therefore, IRIS$^2$ provides additional redundancy compared to Starlink in case of satellite failure (e.g., due to jamming) without compromising coverage, which is particularly desirable for maintaining resilient connectivity regardless of where an emergency occurs.
%TC:ignore
\paragraph{Service continuity}
%TC:endignore
Due to their high orbital velocity, satellites continuously move in and out of visibility with respect to ground UEs. In this case, UEs may need to perform frequent handovers, 
%that is the process of transferring an active UE connection from one satellite (gNB) to another, 
resulting in significant signaling overhead at the Radio Access Network (RAN) and possible service disconnections.
Notably, the visibility period depends on the satellite velocity, which is primarily determined by the satellite altitude.
For a Starlink LEO satellite at 600 km, the speed is 7.56 km/s and the orbital period is approximately 96.7 min, while the actual visibility period is only about 8.5 min. %For an IRIS$^2$ Low-LEO satellite, it further decreases to around 4.6 min, leading to more frequent handovers.
Conversely, IRIS$^2$ MEO satellites move at about 3.07 km/s, and the visibility period increses up to around 6 h.

\begin{comment}
We first evaluate via offline simulations the visibility period for different satellite types, defined as the time during which a UE remains in \acrshort{los} with respect to its serving satellite, and the elevation angle between the two is above a minimum predefined threshold, set to 20$^\circ$ in the simulations.
The visibility period is primarily determined by the satellite altitude, and corresponds to the average time before handover.
%TC:ignore
%\footnote{In this paper we focus on the classical inter-satellite cell-moving handover~\cite{38821}. The study of more advanced handover schemes, such as cell-fixed handover, where the cell is anchored to a fixed point on Earth through dynamic beam steering, rather than to the satellite, as proposed in the 3GPP~\cite{figaro20255g}, is beyond the scope of this work.}
%TC:endignore
For a Starlink LEO satellite at an altitude of 600 km, moving at 7.56 km/s, the orbital period is approximately 96.7 minutes, while the actual visibility period is only about 8.5 minutes. For an IRIS$^2$ Low-LEO satellite, it further decreases to around 4.6 min, leading to more frequent handovers.
Conversely, IRIS$^2$ MEO satellites move at lower orbital speeds (about 3.07 km/s) and shape larger coverage regions on the ground (approximately 2\,500 km), so the visibility period is up to around 360 minutes (i.e., 6 hours).
\end{comment}

These considerations indicate that the MEO satellites in the IRIS$^2$ constellation, which are not available in Starlink, are less exposed to handovers and may improve service continuity, which is particularly critical in emergency scenarios.
%TC:ignore
\begin{figure}[t!]
\centering
\begin{tikzpicture}

\definecolor{darkgray176}{RGB}{176,176,176}
\definecolor{darkorange25512714}{RGB}{255,127,14}
\definecolor{gray128}{RGB}{128,128,128}
\definecolor{lightgray204}{RGB}{204,204,204}
\definecolor{steelblue31119180}{RGB}{31,119,180}

\begin{axis}[
  clip=false,
  legend cell align={left},
  legend style={font=\footnotesize,fill opacity=0.8, draw opacity=1, text opacity=1, draw=lightgray204},
  tick align=outside,
  tick pos=left,
  x grid style={darkgray176},
  xlabel style={at={(axis description cs:0.5,-0.25)},font=\footnotesize},
  tick label style={font=\footnotesize},
  ylabel style={font=\footnotesize},
  xmin=-0.41875, xmax=3.29375,
  xtick style={color=black},
  xtick={0,1,2,3},
  xticklabels={Low-LEO,LEO,MEO, \shortstack{LEO \\ Starlink}},
  xticklabel style={align=center, yshift=0.5ex},
  y grid style={gray128},
  ylabel={Handover rate [UE/s]},
  ymajorgrids,
  ymin=0, ymax=87.5,
  ytick style={color=black}
]
\draw[draw=black,fill=steelblue31119180] (axis cs:-0.25,0) rectangle (axis cs:0,27.7777777777778);
\addlegendimage{ybar,area legend,draw=black,fill=steelblue31119180}
\addlegendentry{$n=10\,000$ UEs}

\draw[draw=black,fill=steelblue31119180] (axis cs:0.75,0) rectangle (axis cs:1,11.9047619047619);
\draw[draw=black,fill=steelblue31119180] (axis cs:1.75,0) rectangle (axis cs:2,0.462320850670365);
\draw[draw=black,fill=steelblue31119180] (axis cs:2.75,0) rectangle (axis cs:3,19.6078431372549);

\draw[draw=black,fill=darkorange25512714] (axis cs:0,0) rectangle (axis cs:0.25,83.3333333333333);
\addlegendimage{ybar,area legend,draw=black,fill=darkorange25512714}
\addlegendentry{$n=30\,000$ UEs}

\draw[draw=black,fill=darkorange25512714] (axis cs:1,0) rectangle (axis cs:1.25,35.7142857142857);
\draw[draw=black,fill=darkorange25512714] (axis cs:2,0) rectangle (axis cs:2.25,1.3869625520111);
\draw[draw=black,fill=darkorange25512714] (axis cs:3,0) rectangle (axis cs:3.25,58.8235294117647);

% parentesi quadra sotto i tick di IRIS^2
\coordinate (L) at (axis cs:0,0);
\coordinate (C) at (axis cs:1,0);
\coordinate (R) at (axis cs:2,0);
\draw[black]
    ([yshift=-16pt]L) -- ([yshift=-20pt]L)
    -- ([yshift=-20pt]C) -- ([yshift=-16pt]C)
    -- ([yshift=-20pt]C)
    -- ([yshift=-20pt]R)
    ([yshift=-20pt]R)
    --([yshift=-16pt]R);
\node[font=\footnotesize] at ([yshift=-30pt]C) {IRIS$^2$};
\end{axis}

\end{tikzpicture}
\caption{Average handover rate for Low-\gls{leo}, \gls{leo}, and \gls{meo} satellites in IRIS$^2$ vs. \gls{leo} in Starlink, as a function of the number of UEs.}
\vspace{-0.5cm}
\label{fig:hr_comparison}
\end{figure}
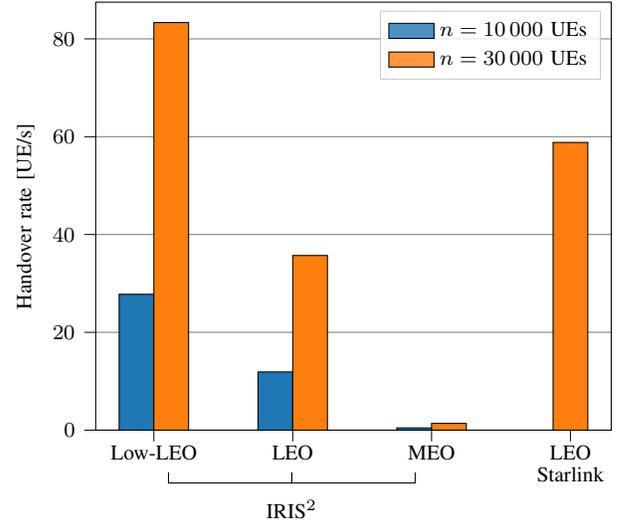
%A fundamental challenge in satellite networks is mobility, which implies that 
%Due to the high-speed mobility of satellites, UEs may need to perform frequent handovers, that is the process of transferring an active UE connection from one satellite (gNB) to another, resulting in significant signaling overhead at the Radio Access Network (RAN) and service continuity challenges.
%In this paper we focus on the classical inter-satellite cell-moving handover~\cite{38821}, in which the cell footprint is related to the serving satellite. The 3GPP is also investigating cell-fixed handover, where the cell is anchored to a fixed point on Earth through dynamic beam steering to reduce the handover frequency~\cite{figaro20255g}, which is beyond our scopes.
%TC:endignore
To prove this claim, in Fig.~\ref{fig:hr_comparison} we plot the handover rate, i.e., the number of UEs performing a handover per second within a cell.
The handover rate increases as the satellite altitude decreases given the shorter visibility period.
At the same time, the cell size also decreases (e.g., the ground radius of a Starlink \gls{leo} cell is $R^{\rm S}_L=600$ km vs. $R^{\rm I}_M=2\,500$ km for an IRIS$^2$ \gls{meo} cell), meaning that more \glspl{ue} may need to handover simultaneously.
Frequent handovers introduce severe signaling overhead if not properly managed, which requires efficient and scalable countermeasures to maintain service continuity.
For IRIS$^2$, the presence of both \gls{leo} and \gls{meo} satellites, despite the lower constellation density, alleviates the impact of handovers, and guarantees more reliable services.
Moreover, since IRIS$^2$ typically serves fewer mission-critical \glspl{ue}, the handover overhead is further reduced. For example, in Fig.~\ref{fig:hr_comparison} the handover rate decreases by around 70\% from $n=30\,000$ to $10\,000$ \glspl{ue}, considering IRIS$^2$ LEO satellites.

Furthermore, we evaluate the capability of IRIS$^2$ to simplify handover operations using MEO satellites.
For Starlink, we consider the conventional cell-moving handover architecture~\cite{figaro20255g}, where all $n$ \glspl{ue} must execute a full handover.
We model each handover event as a three-way handshake whose total delay includes: (i) a propagation component for the transmission of handover signals, which depends on the satellite altitude; and (ii) a processing component $\tau_p$ associated with the processing and execution of handover commands.
In contrast, for IRIS$^2$, we assume that handovers are coordinated at the \gls{meo} satellite~\cite{wei2025iris}, which controls all mobility interactions for the underlying visible LEO satellites. Such handover combines: (i) the propagation delay for the signaling between LEO and MEO satellites; and (ii) the full processing delay $\tau_p$, even though this overhead scales with the number of visible LEO satellites per MEO satellite ($n_L$), rather than with the number of UEs in the cell ($n$), with $n_L \ll n$.
Moreover, a lightweight logical handover is also required between the \gls{ue} and the serving/target LEO satellites, but assumes a reduced processing delay $\alpha\tau_p$, with $\alpha<1$, compared to a full handover procedure~\cite{abdelsadek2021future}.
In Fig.~\ref{fig:handover_overhead} we evaluate the total handover time, i.e., the time required to complete handover for all $n$ \glspl{ue} in a satellite cell, for Starlink and IRIS$^2$.
We set $n_L=28.5$ based on offline simulations, and $\alpha=0.3$, while we vary $\tau_p$ between 0 (ideal) and 30 ms to consider different processing capabilities and implementation configurations, since the 3GPP NR-NTN standard does not currently provide precise handover time specifications. 
We observe that, for $\tau_p<10$ ms, the total handover time of IRIS$^2$ is higher than that of Starlink. This is due to the higher altitude of IRIS$^2$ LEO satellites (around 1\,200 vs. 600 km), which increases the propagation delay of handover signaling. Eventually, as $\tau_p$ increases, the total handover time of IRIS$^2$ decreases by up to 45\% at $n=30\,000$ compared to Starlink, since most of the handover processing is shifted to the MEO satellite, and scales with $n_L\ll n$.
%TC:ignore
\begin{figure}[t!]
    \centering
    \begin{tikzpicture}

\definecolor{darkgray176}{RGB}{176,176,176}
\definecolor{darkorange25512714}{RGB}{255,127,14}
\definecolor{lightgray204}{RGB}{204,204,204}
\definecolor{steelblue31119180}{RGB}{31,119,180}

\begin{axis}[
  tick align=outside,
  tick pos=left,
  x grid style={darkgray176},
  tick label style={font=\footnotesize},
  xlabel style={font=\footnotesize},
  ylabel style={font=\footnotesize},
  xlabel={Handover processing time ($\tau_p$) [ms]},
  % Griglia 0--30 passo 5 al posto della vecchia 1--31 (brutta da leggere).
  % I limiti x NON coincidono con i dati (0--30): sono allargati del 5% per
  % lato (1.5 ms), cioe' il margine di default di matplotlib gia' visibile nel
  % PNG di riferimento. Serve a non far toccare il bordo dell'asse ai tick
  % estremi e ai marker di 0 e 30 ms. enlarge x limits=false evita che
  % pgfplots aggiunga un SECONDO margine sopra a questo.
  xmin=-1.5, xmax=31.5,
  enlarge x limits=false,
  xtick={0,5,10,15,20,25,30},
  xtick style={color=black},
  y grid style={darkgray176},
  ylabel={Total handover time [s]},
  ymajorgrids,
  ymin=0, ymax=1188.1370725477024,
  ytick style={color=black},
  legend cell align={left},
  legend columns=1,
  legend pos=north west,
  legend style={font=\footnotesize, fill opacity=0.8, draw opacity=1, text opacity=1, draw=lightgray204}
]
% Stessa codifica visiva del PNG di riferimento
% (handover_overhead_analysis/plots/handover_overhead_vs_tproc_v4_frac30.png,
% prodotto da plot_handover_overhead_vs_tproc_v4.py): il COLORE codifica il
% sistema (steelblue = Starlink, arancione = IRIS^2 -- lo script usa il verde,
% qui si mantengono i colori originali del repo), mentre lo stile di
% linea/marker codifica n (continuo+cerchio = 10 000, tratteggiato+quadrato
% = 30 000). Ogni voce di legenda nomina esplicitamente sistema e n: NON
% tornare alla legenda fattorizzata su due colonne, che pgfplots riempie per
% righe e che si legge come "10 000 = IRIS^2".
% PROVENIENZA DEI VALORI: data/handover_overhead_vs_tproc_v4.csv campiona
% t_proc su 1,6,...,31 ms, quindi i punti qui sotto NON si leggono dal CSV:
% sono ricalcolati sulla griglia 0,5,...,30 ms con la stessa identica formula
% chiusa di compute_row_v4() (plot_handover_overhead_vs_tproc_v4.py) e le stesse
% costanti di common_params.py, con frac_ue_leo=0.3. Il modello e' lineare in
% t_proc, quindi il cambio di griglia non tocca ne' le pendenze ne' i crossover
% (8.95 ms a n=10 000, 8.70 ms a n=30 000). Bonus: t_proc=10 ms ora cade su un
% punto della griglia, ed e' il valore citato nel testo (152.7 vs 160.0 s a
% n=10 000; 452.9 vs 480.1 s a n=30 000).
% MARKER A t_proc=0: il punto a 0 resta nelle coordinate -- serve a far
% partire la linea dall'origine dell'asse, come nel PNG -- ma il suo marker
% e' soppresso con "mark phase=2" (primo marker disegnato sul 2o punto, cioe'
% 5 ms), perche' un tempo di processing nullo non ha significato fisico.
\addplot[color=steelblue31119180, mark=*, mark phase=2, mark options={fill=steelblue31119180}, line width=1pt] coordinates {(0,60.041537136) (5,110.041537136) (10,160.041537136) (15,210.041537136) (20,260.041537136) (25,310.041537136) (30,360.041537136)};
\addlegendentry{Starlink, $n=10\,000$ UEs}
\addplot[color=darkorange25512714, mark=*, mark phase=2, mark options={fill=darkorange25512714}, line width=1pt] coordinates {(0,122.443643195) (5,137.586159862) (10,152.728676529) (15,167.871193195) (20,183.013709862) (25,198.156226529) (30,213.298743195)};
\addlegendentry{IRIS$^2$, $n=10\,000$ UEs}
\addplot[color=steelblue31119180, mark=square*, mark phase=2, mark options={fill=steelblue31119180}, line width=1pt] coordinates {(0,180.124611407) (5,330.124611407) (10,480.124611407) (15,630.124611407) (20,780.124611407) (25,930.124611407) (30,1080.124611407)};
\addlegendentry{Starlink, $n=30\,000$ UEs}
\addplot[color=darkorange25512714, mark=square*, mark phase=2, mark options={fill=darkorange25512714}, line width=1pt] coordinates {(0,362.609791739) (5,407.752308405) (10,452.894825071) (15,498.037341738) (20,543.179858405) (25,588.322375071) (30,633.464891738)};
\addlegendentry{IRIS$^2$, $n=30\,000$ UEs}
\end{axis}

\end{tikzpicture}
    \caption{Total handover time vs.\ the handover processing delay $\tau_p$ for IRIS$^2$ vs. Starlink, serving $n=10\,000$ and $30\,000$ \glspl{ue}.}
    \label{fig:handover_overhead}
\end{figure}

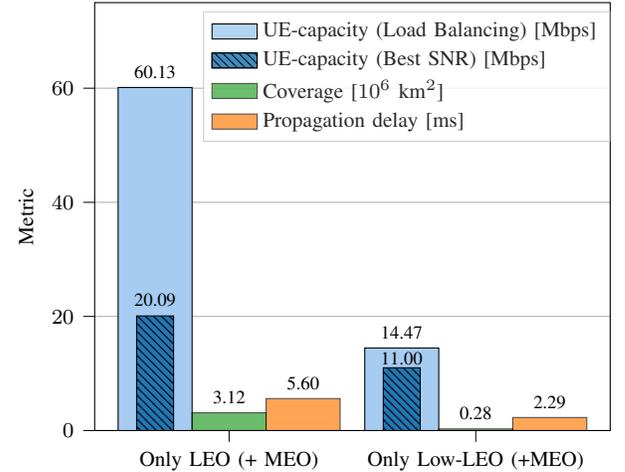
\begin{figure}[t!]
\centering
\begin{tikzpicture}

\definecolor{darkgray176}{RGB}{176,176,176}
\definecolor{color0}{rgb}{0.65,0.80,0.95}
\definecolor{color1}{rgb}{0.172549019607843,0.627450980392157,0.172549019607843}
\definecolor{color2}{rgb}{1,0.498039215686275,0.0549019607843137}
\definecolor{color3}{rgb}{0.12156862745098,0.466666666666667,0.705882352941177}

\begin{axis}[
legend cell align={left},
legend columns=1,
legend style={
  font=\footnotesize,
  fill opacity=0.85,
  draw=white!80!black,
  at={(1,0.98)},
  anchor=north east
},
tick align=outside,
tick pos=left,
tick label style={font=\footnotesize},
ylabel style={font=\footnotesize},
xlabel style={font=\footnotesize},
ylabel={Metric},
xmin=-0.545, xmax=1.545,
xtick style={color=black},
xtick={0,1},
xticklabels={Only LEO (+ MEO), Only Low-LEO (+MEO)},
ymin=0, ymax=75,
ytick style={color=black},
ymajorgrids,
y grid style={darkgray176},
]

% --- bars (unchanged) ---
\draw[draw=black,fill=color0] (axis cs:-0.45,0) rectangle (axis cs:-0.15,60.1259196583913);
\draw[draw=black,fill=color1,opacity=0.7] (axis cs:-0.15,0) rectangle (axis cs:0.15,3.11629472785276);
\draw[draw=black,fill=color2,opacity=0.7] (axis cs:0.15,0) rectangle (axis cs:0.45,5.60325413586761);
\draw[draw=black,fill=color3,postaction={pattern=north west lines}] (axis cs:-0.375,0) rectangle (axis cs:-0.225,20.0869367250674);

\draw[draw=black,fill=color0] (axis cs:0.55,0) rectangle (axis cs:0.85,14.4710874524729);
\draw[draw=black,fill=color1,opacity=0.7] (axis cs:0.85,0) rectangle (axis cs:1.15,0.281978948122647);
\draw[draw=black,fill=color2,opacity=0.7] (axis cs:1.15,0) rectangle (axis cs:1.45,2.2853600013862);
\draw[draw=black,fill=color3,postaction={pattern=north west lines}] (axis cs:0.625,0) rectangle (axis cs:0.775,11.0010886731908);

% --- value labels ---
\node[font=\scriptsize,anchor=south] at (axis cs:-0.3,60.1259196583913) {60.13};
\node[font=\scriptsize,anchor=south] at (axis cs:0,3.11629472785276) {3.12};
\node[font=\scriptsize,anchor=south] at (axis cs:0.3,5.60325413586761) {5.60};
\node[font=\scriptsize,anchor=south] at (axis cs:-0.3,20.0869367250674) {20.09};

\node[font=\scriptsize,anchor=south] at (axis cs:0.7,14.4710874524729) {14.47};
\node[font=\scriptsize,anchor=south] at (axis cs:1,0.281978948122647) {0.28};
\node[font=\scriptsize,anchor=south] at (axis cs:1.3,2.2853600013862) {2.29};
\node[font=\scriptsize,anchor=south] at (axis cs:0.7,9.9) {11.00};

% --- legend ---
\addlegendimage{ybar,area legend,draw=black,fill=color0}
\addlegendentry{UE-capacity (Load Balancing) [Mbps]}
\addlegendimage{ybar,area legend,draw=black,fill=color3,postaction={pattern=north west lines}}
\addlegendentry{UE-capacity (Best SNR) [Mbps]}

\addlegendimage{ybar,area legend,draw=black,fill=color1,opacity=0.7}
\addlegendentry{Coverage [$10^{6}$ km$^2$]}

\addlegendimage{ybar,area legend,draw=black,fill=color2,opacity=0.7}
\addlegendentry{Propagation delay [ms]}

\end{axis}
\end{tikzpicture}
\caption{Downlink per-UE capacity, coverage, and propagation delay considering an IRIS$^2$ constellation with $L^{\rm I}=274$ satellites of the same type (only LEO at 1\,400 km or only Low-LEO at 400 km), besides the MEO satellites. We isolate the performance of the sole LEO and Low-LEO satellites, and set $n=10\,000$. UEs are associated with the satellite that provides the highest per-UE capacity (``Load Balancing'') or SNR (``Best SNR'').}
\vspace{-0.5cm}
\label{fig:throughput_coverage_propagation_scaled}
\end{figure}
\paragraph{Constellation design}
%TC:endignore
Finally, in Fig.~\ref{fig:throughput_coverage_propagation_scaled} we simulate a proposed IRIS$^2$ network with $L^{\rm I}=274$ LEO or Low-LEO satellites, in addition to the \gls{meo} satellites. We set $n=10\,000$, and consider the association models described in Sec.~\ref{sub:association}.
%We compared the per-\gls{ue} capacity, propagation delay, and coverage, isolating the performance of the sole LEO and Low-LEO satellites. 
%Although the altitude difference between standard and Low-\gls{leo} satellites is relatively small, the resulting impact at the network level is not negligible. 
We observe that \gls{leo} satellites, despite the slightly higher propagation delay, provide better coverage than Low-LEO satellites (3.12 vs. 0.28 km$^2$), meaning that the system can serve more UEs, or the same number of \glspl{ue} with fewer satellites, thus reducing deployment and maintenance costs.
Notably, Low-LEO satellites alone may create coverage holes, leaving some \glspl{ue} unconnected.
For these reasons, the per-\gls{ue} capacity of \gls{leo} satellites is up to $4\times$ that of Low-LEO satellites, despite the longer service link. 
Finally, we observe that ``Load Balancing'' improves the per-\gls{ue} capacity by around $3\times$ compared to ``Best SNR'' since satellites can efficiently redistribute \glspl{ue} across different cells based on traffic demands and cell occupancy indicators, rather than congesting the same ``best'' satellite~\cite{10978515}.
%In the latter, all \glspl{ue} associate with the same ``best'' satellite, leading to congestion and potential resource saturation.
%In the former, instead, the satellites can efficiently redistribute \glspl{ue} across different cells based on traffic demands and cell occupancy indicators, which is feasible with second-generation regenerative satellite payloads supporting onboard processing, inter-satellite coordination, and more flexible resource management~\cite{10978515}.
%To mitigate the presence of bottlenecks and fully exploit the available network capacity, 
Although IRIS$^2$ is already in the early stages of deployment, these results suggest that future extensions should prioritize standard \gls{leo} over Low-\gls{leo} satellites to improve coverage and connectivity performance, and support more advanced association models based on load balancing, rather than solely on signal quality.

%TC:ignore
\section{Conclusions}
\label{sec:conclusions}
%TC:endignore
This paper compared Starlink and IRIS\textsuperscript{2} using a unified simulation framework for evaluating them under similar assumptions and working conditions.
Results showed that Starlink outperforms IRIS\textsuperscript{2} in terms of absolute capacity and coverage given its extremely dense constellation, whereas IRIS\textsuperscript{2} also provides desirable features.
First, it generally serves fewer selected mission-critical \glspl{ue},
which can mitigate congestion compared to large-scale commercial systems like Starlink.
Moreover, the presence of MEO satellites in IRIS\textsuperscript{2} guarantees more robust  service continuity, and reduces the handover overhead. 
Finally, we provided design guidelines toward future possible IRIS\textsuperscript{2} extensions, that we claim should be based on \gls{leo} (rather than Low-LEO) satellites, and involve inter-satellite association via load~balancing.

\bibliographystyle{IEEEtran}
\bibliography{bibliography.bib}

\section*{Biographies}

\textsc{Alexander Bonora} [GSM] (alexander.bonora@phd.unipd.it) is a Ph.D. student in Information Engineering at the University of Padua, Italy. He conducts his research within the SIGNET research group at the Department of Information Engineering (DEI). His research interests include 6G wireless networks and non-terrestrial networks, with a particular focus on digital twins, ray tracing, and machine learning.

\textsc{Marco Giordani} [SM] is currently an Associate Professor of telecommunications with the Department of Information Engineering (DEI), University of Padova. His research focuses on 5G/6G cellular networks and PHY/MAC protocol design. He co-authored more than 90 published articles in the area of wireless networks, three of which have received best paper awards. He was a recipient of several awards, including the 2018 IEEE VTS Daniel E. Noble Fellowship Award, the 2021 IEEE ComSoc Outstanding Young Researcher Award for EMEA, the 2024 IEEE ComSoc Young Professional Award for Best Practitioner, and the 2025 IEEE WTC Outstanding Young Researcher Award. He is the Director of the Ph.D. Summer School of Information Engineering (SSIE), the Coordinator of the IEEE Italy Entrepreneurship Committee Local Group, and a member of the same committee for IEEE R8. He serves as anEditor for IEEE Transactions of Wireless Communications and IEEE Transactions on Mobile Computing.

\textsc{Michele Zorzi} [F] (zorzi@dei.unipd.it) received the Laurea and Ph.D. degrees in electrical engineering from the University of Padova. He held academic positions in both Italy and the U.S., and since 2003, he has been a Professor at the Information Engineering Department, University of Padova. His research interests include IoT, non-terrestrial networks, 5G/6G, and underwater communications. He was the EiC of several top IEEE publications, including IEEE Wireless Communication magazine from 2003 to 2005 and IEEE Transactions Communications from 2008 to 2011. He served IEEE Comsoc in several other positions (including a BoG MAL, and the Director of Education and Journals) and received multiple paper and service awards.

%TC:endignore
\end{document}